# MIDAS: SOFTWARE FOR THE DETECTION AND ANALYSIS OF LUNAR IMPACT FLASHES

**José M. Madiedo[1, 2], José L. Ortiz[3], Nicolás Morales[3] and Jesús Cabrera-Caño[2]**

[1] Facultad de Ciencias Experimentales, Universidad de Huelva. 21071 Huelva (Spain).
[2] Facultad de Física, Universidad de Sevilla, Departamento de Física Atómica, Molecular y Nuclear, 41012 Sevilla, Spain.
[3] Instituto de Astrofísica de Andalucía, CSIC, Apt. 3004, Camino Bajo de Huetor 50, 18080 Granada, Spain.

**ABSTRACT**

Since 2009 we are running a project to identify flashes produced by the impact of meteoroids on the surface of the Moon. For this purpose we are employing small telescopes and high-sensitivity CCD video cameras. To automatically identify these events a software package called MIDAS was developed and tested. This package can also perform the photometric analysis of these flashes and estimate the value of the luminous efficiency. Besides, we have implemented in MIDAS a new method to establish which is the likely source of the meteoroids (known meteoroid stream or sporadic background). The main features of this computer program are analyzed here, and some examples of lunar impact events are presented.

**Corresponding author:** José M. Madiedo

Tel.: +34 959219991

Fax: +34 959219983

Email: madiedo@cica.es

**KEYWORDS:** lunar impact, Moon, meteoroids

## 1 INTRODUCTION

According to different estimates, between 40 and 80 tons of interplanetary matter impact Earth every year (Williams and Murad 2002). Most of this is in the form of meteoroids: fragments mostly coming from asteroids and comets and with sizes raging from 100 microns to 10 m. Fireball networks estimate the flux of interplanetary matter impacting our planet by studying the behaviour of meteoroids in the Earth's atmosphere. Thus, different researchers have estimated the impact hazard for our planet by analyzing fireball events simultaneously imaged from several meteor observing stations (see e.g. Ceplecha 2001, Madiedo et al. 2014a).

Meteoroids also impact the Moon, but since this is an airless body, even the smallest meteoroids collide with the lunar surface at high speeds. These violent collisions generate a short duration flash that can be detected from Earth (Ortiz et al. 1999). So the Moon can be used as a giant detector that provides information about meteoroids impacting the Lunar surface. In this way, the impact flux on the Moon can be calculated, and this can be extrapolated to infer the impact flux on Earth (Ortiz et al. 2006, Madiedo et al. 2014b). This method has the advantage that the area covered by one single detection instrument is much larger than the atmospheric volume monitored by meteor detectors employed by fireball networks. However, the results are highly dependent on a parameter called the luminous efficiency, which accounts for the fraction of the kinetic energy of the impactor that is converted into visible light during the impact. Currently this luminous efficiency is not known very accurately, and several estimates have been obtained and employed by different researchers (Bellot Rubio et al. 2000, Ortiz et al. 2002, Swift et al. 2011).

As a continuation of the lunar impacts survey started in 1997 by the second author (Ortiz et al. 1999, 2000), our team is performing since 2009 a monitoring of the night side of the Moon by means of small telescopes and high-sensitivity CCD video cameras (Madiedo et al. 2010, Madiedo et al. 2015). In the framework of this project, which is called MIDAS (Moon Impacts Detection and Analysis System), we have also developed a new software package to identify and analyze lunar impact flashes. Here we describe this software and analyze some results, with special focus on techniques that can be useful to improve the detectability of impact flashes and to estimate the likely source of the meteoroids.

## 2 INSTRUMENTATION AND DATA REDUCTION TECHNIQUES

Since 2009 we are monitoring the night side of the Moon from our observatory in Sevilla, in the south of Spain (latitude: 37.34611 ºN, longitude: 5.98055 ºW, height: 23 m above the sea level). This observatory employs two identical 0.36 m Schmidt-Cassegrain telescopes that image the same area of the Moon, but also two smaller Schmidt-Cassegrain telescopes with diameters of 0.28 and 0.24 m respectively. All of them are manufactured by Celestron. The telescopes are equipped with monochrome high-sensitivity CCD video cameras (model 902H Ultimate, manufactured by Watec Corporation, Japan). These employ a Sony ICX429ALL 1/2" monochrome CCD sensor and produce interlaced analogue imagery according to the CCIR video standard. Thus, images are obtained with a resolution of 720x576 pixels and a frame rate of 25 frames per second (fps). No optical filter is employed. GPS time inserters are used to stamp time information on every video frame with an accuracy of 0.01 seconds. In addition, f/3.3 focal reducers manufactured by Meade are also used in order to increase the lunar surface area monitored by these devices. Major lunar features are easily visible in the earthshine and so these can be used to determine the selenographic coordinates (i.e., latitude and longitude on the lunar surface) of impact flashes. In 2013 we installed an additional telescope at La Hita Astronomical Observatory, in central Spain (latitude: 39.56833 ºN, longitude: 3.18333 ºW, height: 674 m above the sea level). This Newtonian telescope also employs a Watec 902H Ultimate camera and has a diameter of 40 cm.

The data acquisition and reduction pipeline is summarized in Figure 1. As can be seen, the video stream must be digitized in order to analyze the images later on. The video footage is stored on a multimedia hard disk. Then, they are sent to a PC workstation for further processing and analysis. Impact flashes are very short in duration (most of them are contained in just one or two video frames). So, their identification by eye is not practical and computer software is required to automatically detect impact candidates. With this aim we developed the MIDAS software (Moon Impacts Detection and Analysis Software) (Madiedo et al., 2010). This package, which is also employed for data manipulation and reduction of confirmed flashes, is described below.

## 2.1 The MIDAS software

The MIDAS software was mainly developed to process live video streaming and also AVI video files containing images of the night side of the Moon in order to automatically detect flashes produced by the impact of meteoroids on the lunar surface.

One of the advantages of this software is its ability to perform different tasks simultaneously. Thus while the impact flash identification process is in progress, the user can view, edit, and even analyze impact suspects previously detected by MIDAS. The main kernels and features of this package are explained below.

### 2.1.1 Video pre-processing

In general (Figure 1), the source AVI video files need to be pre-processed before performing the flashes identification procedure. For instance, since our cameras generate interlaced video it is convenient to de-interlace it. In this way, typical undesired artifacts produced in the images by the video interlacing technology are removed. Video deinterlacing is also particularly useful when the photometric analysis of impact flashes is performed, as explained below. During the deinterlacing process, the software extracts both interlaced video fields (top and bottom) and generates another AVI file where these fields are included as independent frames placed one after the other. This automatically doubles the frame rate of the original video rate. In our case, since CCIR cameras are employed, we go from 25 fps to 50 fps.

Video pre-processing also addresses some issues that may interfere with the impact flash detection process and, specifically, with the detectability of fainter flashes. With this aim, over 20 image and video processing filters have been implemented in MIDAS. Thus, for instance, the significant amplification of signal of our CCD video cameras (up to 60 dB according to the manufacturer's specifications) can give rise to a large number of false detections, and different approaches can be employed to address this issue. For instance, images can be smoothed by MIDAS by means of a median filter, which averages every pixel according to the brightness of the surrounding pixels within a given user-specified radius. One of the drawbacks of this method is that it gives rise to blurred images. Besides, it poses some difficulties to the detectability of fainter flashes, since these events appear in very small pixel clusters whose brightness is significantly reduced in the resulting images when these are averaged with the less-luminous surrounding pixels. Video reduction to 2:1 size is an alternative useful technique implemented in MIDAS for image smoothing (Cudnik 2009). According to this technique, both the width and height of each video frame are halved. When resampling the video to the new size, the intensity of each pixel in the output frames is calculated by applying a cubic spline interpolation filter to the intensity of the closest 16 source neighbour pixels (Keys 1981). The main benefits of video reduction to 2:1 size are that image noise is decreased and that the video processing time is significantly reduced: the total number of pixels to monitor and so also the processing time decreases by a factor of 4. When employing this video size reduction technique it should be taken into account that, since impact flashes cover more than one single pixel in the images, these features do not disappear during the 2:1 reduction process and so these can still be easily identified in the resulting video. However, the so-called temporal smoothing has proven to be a more efficient method to deal with false positives arising from noisy images. This method, which produces sharper images, applies a transformation filter to the video file, so that each frame is substituted by a weighted composition of m frames around it. In particular, the temporal smoothing technique implemented in MIDAS employs the following equation:

$$I_n^{out}(x,y) = \frac{\sum_{i=n-m}^{n+m} I_i^{in}(x,y)\cdot W(i)}{\sum_{i=n-m}^{n+m} W(i)} \qquad (1)$$

where $I_n(x,y)$ is the intensity (pixel value, ADU) for the pixel located at the position (x,y) in frame n. The weighting function W employed by MIDAS is given by

$$W(i) = \max\left\{0, \alpha - \frac{\left|I_i^{in}(x,y) - I_n^{in}(x,y)\right|}{2^\beta}\right\} \qquad (2)$$

In Eq. (2), $\alpha$ and $\beta$ are, together with parameter m in Eq. (1), the free user-selectable parameters of the temporal smoothing transform. The size of the window around each video frame (i.e., the number of frames taken into account to smooth each frame) is given by the parameter m. As can be seen, Eq. (2) assigns a larger weight to the central frame in this window and yields smaller contributions for adjacent frames. However, the larger is $\beta$, the stronger is the contribution of adjacent frames with respect to the central one. Another consequence of Eq. (2) is that if the pixel luminosity of any adjacent frame with respect to the central one is too large (so that W(i) yields a negative value), the contribution of this adjacent frame is zero since W(i) must be always set to 0 in that case. Then, it is easy to notice that the parameter α is related to the maximum allowed intensity difference between these frames. MIDAS sets the value of this parameter equal to the average noise level in the images. In this way, differences due to this noise are smoothed, but any difference arising from the sudden appearance of an impact flash will remain enhanced. This is critical to improving the detectability of dimmer flashes.

### 2.1.2 Mask definition

Some regions in the images should be ignored during the impact flash identification process in order to prevent false detections. Thus, as is described in more detail below, the procedure used by MIDAS to detect impact flashes is based on the comparison of consecutive video frames to identify sudden pixel brightness changes that could be indicative of the collision of meteoroids on the lunar surface. This, however, implies that other phenomena, such as for instance star scintillation, can give rise to false detections. So, a mask must be defined to indicate that regions on the image containing stars should be ignored and just the area covered by the lunar disk should be considered. In this context, a mask is an image with the same size (width and height) as the video frames to be analyzed that indicates, by means of a colour code, which areas are going to be taken into account during the detection process and which of them will be ignored. For MIDAS, masked areas are indicated with a red colour. Additional regions to be ignored include also those with fast changing features that could give rise to a high number of false detections. This is the case of the region in the image containing the information stamped by the GPS time inserter.

A mask editor has been implemented in the software to define which areas should be ignored during the flash detection process. Using the mouse the user marks the regions that will be ignored. The editor provides an option to mask the regions located outside the lunar disk. In this case, by clicking on at least 3 different points located at the edge of the Moon, the software calculates the size of the Moon in the image and masks the corresponding areas. This, obviously, excludes stars and so prevents any issues resulting from star scintillation.

#### 2.1.3 Moon disk calibration

In general, the high sensitivity of the Watec Ultimate cameras makes lunar features easily identifiable. These can be used to calibrate the lunar disk, so that when an impact flash is detected we can automatically convert x, y coordinates in the images into selenographic (latitude and longitude) coordinates. The software performs this calibration by means of an implemented tool that takes into account the known position of at least three different features on the Moon's surface. Then, if an event is detected, the software automatically provides its (x,y) coordinates but also the corresponding values of latitude and longitude on the lunar surface. This calibration is optional. When it is not performed, just (x,y) coordinates are provided by the software.

### 2.1.4 Automated identification of impact flashes

The impact flash identification routines are the main kernel of the MIDAS software. Although these can analyze live imagery generated by our CCD video cameras, it is advisable to first digitize these images as AVI video files and process these files later on. In this way, these videos can be pre-processed to address some issues that may interfere with the impact flash detection process, as explained above. Additional improvements are being currently made to handle also MPEG video files, since this is the native format of some video recorders. Although uncompressed AVI video files are preferred in order to get better image quality, the software also processes compressed AVI files provided that the corresponding compression codec is properly installed and configured. According to the tests we have performed, XVID and DIVX compression formats with high quality compression settings have, for instance, provided very good results with MIDAS.

In order to indentify a potential impact flash, the software compares consecutive video frames and detects clusters of pixels showing brightness changes that exceed a given (user defined) threshold value. The selected detection settings are fundamental in order to get optimal results. These parameters depend, among other things, on the configuration of the imaging telescope and camera. The user must specify the minimum cluster size and brightness threshold that can give rise to a potential detection. Events detected in this way must be compared later on with impact flash candidates identified by other telescopes within the same network in order to check which of them are really produced by the impact of meteoroids on the lunar surface.

When a flash is detected, the software creates a small AVI video file with the corresponding images and stores it on a database (see below) together with some basic information which includes appearance time and position. This file is small in terms of the short duration of the equivalent video clip, since just the first frame where the flash is identified and also X frames later and before are included in it. Since impact flashes can be very faint, these are framed with a red square within this small AVI file to allow for a better location and visualization. The list of video files generated in this way can be browsed even during the detection process, and events can be viewed, edited and even removed if necessary. No brightness information is obtained at this stage, since the photometric analysis is performed later on.

Impact flash detection routines have been tested and validated in different ways. Thus, for instance, MIDAS was tested by repeating the analysis of the imagery obtained by Ortiz et al. (2005) during the monitoring of Leonid lunar impact flashes in 2004. On the other hand, a software called Impact Flash Simulator was also developed in the framework of the MIDAS project for testing purposes. This simulator creates artificial impact flashes and inserts them on real footage recorded by the lunar impacts monitoring system. Thus, this tool produces new videos where these artificial flashes are created with the desired duration and brightness, and at a given position. These testing videos were used to determine the optimal detection settings to guarantee that the MIDAS software is able to identify similar events in real videos.

### 2.1.5 Impacts database

For every telescope in the system, MIDAS produces a database containing impact flash candidates. Appearance time and position are stored together with a small AVI video frame where the event is framed, as described above. However, as has been previously mentioned, other phenomena can give rise to false detections. Impact flash candidates must therefore be compared with the detections obtained with at least one more telescope in order to confirm if they are in fact produced by meteoroids impacting the lunar surface. MIDAS automatically performs this procedure by comparing the databases created for each telescope. As a result of this comparison, the software creates a list containing the different events, where confirmed impact flashes and false positives are indicated.

These databases can be browsed and edited. So, data related to a given event can be examined or even removed in the case of false detections. If necessary, further analysis can be performed on these events such as, for instance, the photometric analysis described below.

**2.1.6 Photometry**

The estimation of the kinetic energy of the impactor is of paramount importance in order to estimate, for instance, the diameter of craters resulting from lunar impacts, but also to establish the impact flux on the Moon and on our planet. To infer the value of this kinetic energy, the energy radiated as visible light by the flash must be obtained by analyzing the light curve of the event (Ortiz et al. 2006, Madiedo et al. 2014b). With this aim, a software module for photometric analysis has been implemented in MIDAS. This tool calculates the evolution with time of impact flashes brightness. Usually this routine is fed with the AVI video files automatically generated by MIDAS during the flashes detection process. However, the original AVI video file recorded by the camera connected to the telescope can be also used when necessary.

In a first step the software obtains brightness versus time data for the flash, where luminosity is expressed in instrument units (i.e., pixel value, which in the case of 8-bit analog to digital conversion ranges from 0 to 255). This is also done for a series of reference stars selected by the user. Next, since the magnitude of these calibration stars are known, a magnitude versus time plot can be obtained for the impact flash. The photometric analysis routine employed by the MIDAS software has been successfully tested and validated with the results obtained with the Limovie software (Miyashita et al. 2006). As commented previously, the video files containing the impact flash are deinterlaced before this photometric analysis is performed, effectively doubling the frame rate. This is very useful since impact flashes are in general very short in duration, and this procedure doubles the temporal resolution of the light curve.

Once the flash magnitude is obtained, the radiated power can be calculated from the equation

$$P = 3.75 \cdot 10^{-8} \cdot 10^{(-m/2.5)} f \pi \Delta\lambda R^2 \qquad (3)$$

where P is the radiated power (in watts), m is the time-dependent magnitude of the flash, $\Delta\lambda$ is the width of the filter passband, R is the Earth-Moon distance and f is a dimensionless factor that describes the degree of anisotropy of light emission. In the equation, the factor $3.75\ 10^{-8}$ is the flux density in W $m^{-2}$ $\mu m^{-1}$ for a magnitude 0 source according to the values given in Bessel (1979). For events where light is isotropically emitted from the surface of the Moon f = 2, while f = 4 if light is emitted from a very high altitude above the lunar surface.

The energy released as visible light on the Moon ($E_r$) is calculated by MIDAS by performing a numerical integration of the radiated power P with respect to time. This magnitude is related to the kinetic energy E of the impactor by means of the relationship $E_r = \eta E$, where $\eta$ is the luminous efficiency.

### 2.1.7. Meteoroid source

The association of a lunar impact flash with a given meteoroid stream is fundamental in order to estimate, for instance, the velocity of the impactor and its mass, or the size of the crater produced by the collision. Linking a meteoroid with a meteoroid stream is straightforward when meteors produced by the ablation in the atmosphere of these particles of interplanetary matter are recorded. Thus, provided that a meteor is simultaneously registered from at least two different sites, the radiant can be easily determined and the heliocentric orbit of the meteoroid can be calculated (Ceplecha 1987). Once these orbital elements are known, dissimilarity criteria can be employed to test the potential association between the inferred heliocentric orbit of the meteoroid and that of the meteoroid stream (see, e.g., Williams 2011, Madiedo et al. 2013). For the calculation of the orbit of the meteoroid the knowledge of the velocity vector is fundamental. For meteoroid impacts taking place on the lunar surface the velocity vector is unknown, since just the impact position is available from observations. So, the above described approach cannot be employed and, in fact, in this case it is not possible to unambiguously associate an impact flash with a given meteoroid stream. Nevertheless, a likely association can be inferred from a statistical point of view. Thus, we have defined a parameter p to measure the probability of this association, and the code to estimate its value has been implemented in MIDAS. This parameter, which ranges from 0 to 1, is given as the quotient between the number N of impacts per unit time that can be produced by that meteoroid stream at the location where the flash was detected and the total number of impacts that can be produced by all available sources at the same location. For simplicity we will first consider that these alternative sources include only sporadic events:

$$p = \frac{N^{ST}}{N^{ST} + N^{SPO}} \quad (4)$$

According to this definition, the lower the value of p, the higher the probability that the impact flash is produced by a sporadic meteoroid. On the contrary, the closer to 1 is p, the higher the probability that the event is associated to a particle belonging to the meteoroid stream considered in this analysis. It is obvious that $N^{ST} = 0$ for those flashes located outside the area on the Moon that can be impacted by meteoroids from that stream. For events within that area, $N^{ST}$ is related to the ZHR of members from this stream impacting the Moon:

$$N^{ST} = \varepsilon \cdot ZHR^{ST}_{Moon} \quad (5)$$

where ε is a geometric factor that will be defined below. $ZHR^{ST}_{Moon}$ can be related to the ZHR of this stream on Earth ($ZHR^{ST}_{Earth}$) by means of this relationship:

$$ZHR^{ST}_{Moon} = \sigma \cdot \gamma \cdot ZHR^{ST}_{Earth} \qquad (6)$$

In addition, the ZHR on Earth at solar longitude λ (which corresponds to the time of detection of the impact flash) can be related to the peak ZHR by means of (Jenniskens 1994):

$$ZHR^{ST}_{Earth} = ZHR^{ST}_{Earth}(max) \cdot 10^{-b|\lambda-\lambda_{max}|} \qquad (7)$$

where $ZHR^{ST}_{Earth}(max)$ is the peak ZHR on Earth (corresponding to the date given by the solar longitude $\lambda_{max}$). The values for the peak ZHR for different meteoroid streams and the corresponding solar longitudes for these maxima can be obtained, for instance, from Jenniskens (2006). For streams with non-symmetrical ascending and descending activity profiles or with several maxima, Eq. (7) should be modified according to the expressions given in Jenniskens (1994). The factor σ accounts for the fact that, in general, the meteoroid stream will be at a different distance from the Earth than from the Moon and, so, the density of meteoroids from that stream on both bodies would be different. If we assume a simple situation where this stream can be approximated as a tube where the meteoroid density decreases linearly from its central axis, the following definition can be adopted for σ:

$$\sigma = d_{Earth}/d_{Moon} \qquad (8)$$

where $d_{Earth}$ and $d_{Moon}$ are the distance from the center of the meteoric tube to the Earth and the Moon, respectively.

Besides, since the gravitational field of both bodies is very different, a factor γ has been introduced to account for the different gravitational focusing effect for meteoroids on Earth and Moon. The gravitational focusing factor Φ is given by

$$\Phi = \left(1 + \frac{V_{esc}^2}{V^2}\right) \qquad (9)$$

where $V_{esc}$ is the escape velocity of the central body and V the meteoroid velocity. The factor γ is then obtained by dividing the focusing factors Φ corresponding to the Moon and the Earth. For sporadic meteoroids this velocity-dependent focusing effect is higher for the Earth by a factor of 1.3 (Ortiz et al. 2006), and so $\gamma^{SPO} = 0.77$. The value of γ for a given stream will be labelled as $\gamma^{ST}$.

The parameter ε in Eq. (5) can be obtained by analyzing how particles in the meteoroid stream intersecting the Moon are spread over the lunar surface. The stream will be considered as cylindrical in shape. Thus, without loss of generality we will consider a great circle on the Moon passing through the location of the subradiant point on the lunar surface (Figure 2). In this two-dimensional problem we will define a (linear) density of meteoroids in the stream ρ, which will be considered for simplicity as constant in this scale. The modification for a non constant density is straightforward. Besides, in this analysis the Moon will be considered as spherical, with a radius R. From these assumptions and on the basis of purely geometrical grounds it can be easily shown that the density of meteoroids impacting on the lunar surface ρ' at a given point located at an angular distance φ from the subradiant point is not constant, since it depends on the position angle φ:

$$\rho' = \rho \cdot \cos(\varphi) \quad (10)$$

Thus, the geometrical factor $\varepsilon=\cos(\varphi)$ is responsible for the dependence of $N^{ST}$ on the position on the lunar surface as expressed in Eq. (5).

We will assume that sporadic meteoroids hit the Earth-Moon system isotropically. Since sporadic events come from diffuse sources, these meteoroids may impact at any position on the lunar surface. Also for this reason we will employ hourly rates (HR) instead of zenithal hourly rates (ZHR) for sporadic meteoroids, since the dependence of the hourly rate on the average zenithal radiant distance would be very weak (Dubietis and Artl 2010). We will also take into account the different gravitational focusing effect between Earth and Moon for sporadic meteoroids by means of a factor γ. However, we will consider that the number per unit time of sporadic meteoroids is not affected by any distance factor σ:

$$N^{SPO} = HR^{SPO}_{Moon} = \gamma^{SPO} \cdot HR^{SPO}_{Earth} \qquad (11)$$

So, we finally have the following expression for p:

$$p = \frac{\gamma^{ST}\cos(\varphi)\cdot\sigma\cdot ZHR^{ST}_{Earth}(max)10^{-b|\lambda-\lambda_{max}|}}{\gamma^{SPO}HR^{SPO}_{Earth} + \gamma^{ST}\cos(\varphi)\cdot\sigma\cdot ZHR^{ST}_{Earth}(max)10^{-b|\lambda-\lambda_{max}|}} \qquad (12)$$

If at the time of detection of the impact flash n additional meteoroid streams with significant contribution to the impact rate (and with compatible impact geometry) must be considered, the denominator in Eq. (12) must be modified to account for this contribution:

$$p = \frac{\gamma^{ST}\cos(\varphi)\cdot\sigma\cdot ZHR^{ST}_{Earth}(max)10^{-b|\lambda-\lambda_{max}|}}{\gamma^{SPO}HR^{SPO}_{Earth} + \gamma^{ST}\cos(\varphi)\cdot\sigma\cdot ZHR^{ST}_{Earth}(max)10^{-b|\lambda-\lambda_{max}|} + \kappa} \qquad (13)$$

where

$$\kappa = \sum_{i=1}^{n}\gamma_i^{ST}\cos(\varphi_i)\sigma_i ZHR^{ST}_{i,\ Earth}(max)10^{-b_i|\lambda-\lambda_{i,max}|} \qquad (14)$$

accounts for these n additional streams.

For the average hourly rate of sporadic events we have $HR^{SPO}_{Earth}$ =10 meteors $h^{-1}$ (Dubietis and Artl 2010).

Although the relationships (12) and (13) derived above from Eq. (4) would in principle provide a simple rule to associate impact flashes with meteoroid streams, their major drawback is that they do not take into account the fact that only those meteoroids capable of producing detectable impact flashes from Earth should be included in the computations. Thus, in order to perform a more precise analysis only those impacts with a kinetic energy above a threshold value $E_m$ capable of producing a detectable flash from Earth should be considered, instead of taking into account the total number N of meteoroids impacting per unit time. In other words, the mass distribution and the impact velocity V also plays an important role in this computation. For instance, slow meteoroid streams with most meteoroids having small masses would produce a lower number of detectable impact flashes than faster streams containing more massive members, even if the ZHR of the latter is lower. According to Eq. (2) in Bellot Rubio et al. (2000), the following multiplicative factor should be included in both the numerator and the denominator of Eq. (4) to include in the analysis only those events with a kinetic energy above $E_m$:

$$\nu=\left[\frac{m_o V^2}{2}\right]^{s-1} E_m^{1-s} \qquad (15)$$

where V is the impact velocity, $m_o$ is the mass of a shower meteoroid producing on Earth a meteor of magnitude +6.5 and s is the mass index, which is related to the population index r (the ratio of the number of meteors with magnitude m+1 or less to the number of meteors with magnitude m or less) by means of the relationship

$$s=1+2.5\log(r) \qquad (16)$$

In this way, Eq. (12) should be modified in the following way:

$$p=\frac{\nu\cdot\gamma^{ST}\cos(\varphi)\cdot\sigma\cdot ZHR_{Earth}^{ST}(\max)10^{-b|\lambda-\lambda_{max}|}}{\nu^{SPO}\cdot\gamma^{SPO}HR_{Earth}^{SPO}+\nu\cdot\gamma^{ST}\cos(\varphi)\cdot\sigma\cdot ZHR_{Earth}^{ST}(\max)10^{-b|\lambda-\lambda_{max}|}} \qquad (17)$$

If this correction for the minimum detectable energy is not performed, one effectively assumes that impact flashes produced by meteoroids with a mass $m_o$ (the mass capable of producing a mag. 6.5 meteor in the Earth's atmosphere) are detectable. This mass, however, is several orders of magnitude below the mass of the meteoroids producing the dimmest impact flashes currently detectable by employing small telescopes.

For a most general case, where more than one meteoroid stream is considered, Eq. (13) should be also modified accordingly. A different value of $\nu$ should be employed for each stream, since from Eq. (15) it is obvious that this parameter depends on both the mass index s and the mass $m_o$. The value of $m_o$ for different streams can be obtained from equations (1) and (2) in Hughes (1987). The maximum visual magnitude for detectable impacts ($m_{max}$) would depend, among other factors, on the experimental setup employed. This maximum magnitude would correspond to the minimum radiated energy $E_{rm}$ detectable from observations on Earth, which in turn would be related to the minimum kinetic energy $E_m$ by means of the luminous efficiency:

$$E_{rm}=\eta E_m \qquad (18)$$

$E_{mr}$ can be easily calculated from the radiated power measured on Earth, by assuming a typical value for $m_{max}$ ~ mag. 10 and by using the equations given in the previous section. It is worth noting that the minimum detectable power P on Earth, which is given by Eq. (3), depends on the value of the Earth-Moon distance R. Thus, as expected, the minimum radiated power (and hence the minimum impactor kinetic energy) necessary to give rise to a detectable impact on Earth is larger for larger R values.

The behaviour of the probability parameter p defined by Eq. (17) can be understood through Figures 3 to 5, which have been obtained by setting $\varphi=45^o$ (the value of the most probable impact angle) and $\sigma=1$. Thus, as Figure 3 shows, when the values of the population index r and the luminous efficiency $\eta$ are fixed, the probability p decreases as the geocentric velocity of the meteoroid stream increases. This effect is more important the lower is the zenithal hourly rate of the meteoroid stream. The decrease in p as $V_g$ increases accounts for the fact that the larger the meteoroid velocity, the smaller $m_o$ (the mass of meteoroids producing mag. 6.5 meteors on Earth). This means that, if for instance, the ZHR of the stream coincides with the HR of sporadics, the meteoroid stream would be mainly populated by less massive particles, and so the frequency of larger meteoroids in this stream with high enough kinetic energy to produce a detectable lunar impact flash is smaller. To compensate for this effect, the ZHR of shower meteoroids would need to exceed the HR of the sporadic background, and this difference should be larger the higher is the velocity of these shower meteoroids. Against this explanation one could argue that even if $m_o$ is smaller when $V_g$ increases, this higher $V_g$ value would still keep the kinetic energy of shower meteoroids high enough, and so this would not have a negative effect of the detectability of impact flashes produced by the meteoroid stream. However, simple computations reveal that the decrease in $m_o$ predicted from Eqs. (1) and (2) in Hughes (1987) is so strong that the increase in velocity cannot compensate for the loss of kinetic energy produced by this decrease in mass. The effect on p of the population index r for fixed stream ZHR and luminous efficiency is shown in Figure 4. Here, larger r values give rise to lower probabilities for the corresponding stream, and this effect is larger for larger meteoroid velocities. This is the expected behaviour, since the larger r leads to a higher number of less massive particles in the meteoroid stream. Consequently, the frequency of meteoroids of sufficiently high kinetic energy becomes lower. Finally, Figure 5 shows the effect of the luminous efficiency on the probability parameter for fixed meteoroid velocity and ZHR. By definition, the higher is $\eta$ the larger is the fraction of the kinetic energy of the impactor that is converted into visible light. So, the minimum kinetic energy necessary to produce a detectable flash is lower. This, as shown in Figure 5, implies that for streams with a population index lower than the population index considered for the sporadic background ($r = 3.0$), the values of p are higher, since the stream is in that case populated with larger meteoroids and so it is more likely to find particles capable to exceed this energy threshold and produce detectable flashes. On the contrary, for streams with $r > 3$ the situation is the inverse, since in that case the stream is populated with smaller meteoroids.

#### 2.1.8 Additional tools

Several additional tools have been implemented in MIDAS to plan the lunar monitoring sessions and also as add-ons for data reduction. One of these tools is the so-called lunar calendar, which consists of an interactive window that displays the lunar phase for every day of the selected month, together with the corresponding illuminated fraction of the lunar disk. This is helpful to visualize which dates are suitable for impact flash monitoring (those with an illuminated fraction of the lunar disk ranging between 0.1 and 0.6). By clicking on a given day the software provides additional information, such as the evolution with time of the Moon's position (altitude and azimuth from the observing location) and also local moonrise and moonset times.

MIDAS also determines which meteoroid radiants are active on a given date and which portion of the lunar disk should receive the impact of meteoroids belonging to these swarms. We employ this tool to plan which dates and/or areas on the lunar surface are more interesting to monitor. To obtain this information the software also includes a comprehensive database of meteoroid streams and their main features (activity period, velocity, etc.). These data have been taken from the IAU Meteor Data Center website (http://www.astro.amu.edu.pl/~jopek/MDC2007/). This database is open and interactive. Thus, the user may browse it to view or modify the characteristics of a given stream and new meteoroid streams can even be added. This database is also used by MIDAS for other purposes as, for instance, to determine if an impact flash can be associated to a given meteoroid stream. It is also used by the so-called "Radiant Observing planner", which is a tool that calculates which years are more convenient to observe flashes produced by meteoroids coming from a given meteoroid stream, by taking into account the lunar phase and geometry.

Another tool implemented in MIDAS allows calculating the total area monitored by a given telescope just from one image taken by the system. To perform this calculation the images should contain at least a part of the Moon's limb. The software uses this limb and its curvature to infer the radius of the lunar disk in pixel units. By scaling this value with the well-known Moon's radius in km, and by taking into account which is the portion of the Moon appearing in the images, the value of the monitored area in $km^2$ is obtained.

### 2.1.9. Detection of impacts on other bodies in the Solar System

Although the MIDAS software was designed to identify impact flashes on the lunar surface, we have also considered the possibility of using it to detect impact flashes on other bodies in the Solar System. The first direct observation of one of these collisions took place in July 1994, when Comet D/1993 F2 (Shoemaker-Levy 9) impacted Jupiter. The interest for these events increased a few years ago as a result of the casual detection of several impact flashes on this planet by amateur astronomers (Orton et al. 2011; Fletcher et al. 2010, 2011; de Pater et al. 2010). This led our team to design in September 2010 a campaign with the 1.25 m telescope at the Calar Alto Astronomical Observatory (Spain) with the aim to optimize and test techniques for the detection of these impacts. Despite not recording any impact on this planet during this campaign, this allowed us to implement in the MIDAS software new procedures to optimize the detection of flashes on Jupiter and other planets. Some of the differences with respect to the detection of impact flashes on the Moon were related to the fact that the influence of seeing was much more critical, since much higher magnifications must be employed. In addition, the monitoring of Jupiter implies dealing with a much more dynamic environment. Thus, for instance,

Galilean moons often produce transits and shadows on the planet. This must be taken into account when, for instance, the image mask is defined.

## 3 SELECTED RESULTS

As an example of the methods and techniques employed by the MIDAS software we have selected a series of impact flashes recorded in the framework of our lunar monitoring project in 2011 and 2012. These flashes are listed in Table 1. Among these, discussed below, there are flashes associated to sporadic sources, but also to several meteoroid streams. Their apparent magnitude ranged from 8.0 to 9.8.

Flashes in Table 1 were simultaneously recorded by two of the telescopes operating at our observatory in Sevilla: one of the 0.36 m SC telescopes and the 0.28 m SC system. It must be taken into account that only the simultaneous detection from two different sites can discard the possibility of satellite glints. Since these events were detected at a single observatory, this possibility cannot be discarded completely.

During the activity period of major meteor showers the observing sessions were planned with the aid of the MIDAS software. In this way we could determine which areas on the lunar disk should be monitored in order to increase the possibility of flashes detection. On the contrary, the telescopes were oriented to an arbitrary region in order to cover a common maximum area. In any event, as usual, the terminator was always avoided in order to prevent the negative effects of light coming from the diurnal surface of the Moon. The calculated extension of the monitored area ranged between $(4.1 \pm 0.4)\cdot 10^6$ and $(5.6 \pm 0.5)\cdot 10^6$ $km^2$ for the 0.36 m telescope and between $(7.4 \pm 0.7)\cdot 10^6$ and $(7.6 \pm 0.7)\cdot 10^6$ $km^2$ for the 0.28 m telescope.

To identify these impact flashes in the imagery generated by the cameras, different detection procedures were employed. On a first run, the images were processed by the MIDAS software without employing any of the image smoothing techniques described above. This method is very fast, since the processing time is comparable to the actual duration of the recorded footage when the video is analyzed on a Pentium IV PC computer running at 3 GHz. However, in general it also generates a large list of false positives. By following this procedure, MIDAS identified all of the events listed in Table 1 on both telescopes, except for the mag. 9.8 flash recorded on 27 March 2012. In fact, according to our experience, the detectability limit in non-smoothed images corresponds to a visual magnitude of about 9, although this depends also on the specific observing conditions. This fainter flash was identified on a second run, where the source video files were previously transformed by means of the temporal resolution filter defined by Eqs. (1) and (2). We employed the optimal parameters found for this filter according to our experience. Thus, the smoothing window size was set with $m = 4$, with a filter strength $\beta = 4$. The value of $\alpha$ was set to 20.

## 3.1. Meteoroid source

Most of the flashes listed in Table 1 were not detected within the activity period of any noticeable meteor shower and so they have been associated to sporadic sources. However, events recorded on 26 February 2012 and 27 March 2012 took place close to the maximum activity peak of the δ-Leonids (DLE) and the Virginids (VIR), respectively. On the other hand, the flash registered on 26 July 2012 might have been produced by an α-Capricornid (CAP) meteoroid, and two events recorded on October 20, 2012 could in principle be associated to the Orionids (ORI). As shown in Figure 6, the impact geometry was favourable and the selenographic coordinates of the flashes are compatible with these meteoroid streams. The average geocentric velocity $V_g$ for meteoroids in these streams is 29 km $s^{-1}$ for the δ-Leonids, 30 km $s^{-1}$ for the Virginids, 25 km $s^{-1}$ for the α-Capricornids and 67 km $s^{-1}$ for the Orionids (Jenniskens 2006). To validate these potential associations, the probability parameter p defined by Eq. (17) was calculated. We have assumed $HR^{SPO}_{Earth}$=10 meteors $h^{-1}$ (Dubietis and Artl 2010). For simplicity, we have considered $\sigma = 1$. To calculate $\nu^{SPO}$ from Eq. (15), we have adopted $r^{SPO} = 3.0$ (see e.g. Dubietis and Artl 2010; Rendtel 2006), which according to Eq. (16) implies that $s^{SPO} = 2.2$. The value of $m_o$ for sporadic meteoroids obtained from Eqs. (1) and (2) in Hughes (1987) yields $m_o$=5.0·$10^{-6}$ kg for an average geocentric velocity of around 20 km $s^{-1}$ (Brown et al. 2002). The impact velocity of these sporadic particles has been set to 17 km $s^{-1}$ (Ortiz et al. 1999). As mentioned before, the value of $E_m$ (by assuming a maximum visual magnitude for detectable impacts of about 10) obtained from Eq. (3) will depend on the Earth-Moon distance and so its value will be different for different dates. Hence, $\nu^{SPO}$ will also vary. The calculated values for these quantities are listed in Table 2. By proceeding in the same way we have calculated the value of ν for the five flashes listed in Table 2. In this table, the values of the zenithal hourly rate, impact angle, geocentric velocity and population index are set by assuming that these events are associated to the meteoroid streams specified in the last column. The impact velocity V, obtained from the geocentric velocity $V_g$ by the method described in Madiedo et al. (2014b), is also shown in this Table. The assumed value of the luminous efficiency was $\eta = 2\cdot10^{-3}$, which is the value determined for Leonid lunar impact flashes in e.g. Bellot Rubio et al. (2000) and Ortiz et al. (2002), and also used by Ortiz et al. (2006) to determine impact fluxes on Earth. This value is close to the $\eta=1.5\cdot10^{-3}$ value used by other investigators (see e.g. Swift et al. 2011). The gravitational focusing effect can be neglected, since $\gamma^{ST}/\gamma^{SPO}$ was found to be very close to 1.

It is worth noting that, except for the event potentially associated with the α-Capricornids, the time of recording of the flashes considered here was less than one day (i.e., less than 1 degree in solar longitude) to the maximum activity peak of their respective suspected radiants. So, for the shower activity term in the numerator of Eq. (17) we have

$$ZHR^{ST}_{Earth} = ZHR^{ST}_{Earth}(\max)\cdot10^{-b|\lambda-\lambda_{\max}|} \approx ZHR^{ST}_{Earth}(\max) \qquad (19)$$

For the α-Capricornids, which have a peak activity around July 30, our meteor observing stations located in Spain (Madiedo & Trigo-Rodríguez 2008, Madiedo 2014) measured on 26 July 2012 a zenithal hourly rate of about 6 meteors $h^{-1}$. In this way p yields 0.13 for the flash recorded on 26 Feb. 2012, 0.27 for the event imaged on 27 March 2012, 0.82 for the flash registered on 26 July 2012, 0.19 for the event recorded on 20 Oct. 2012 at 20h05m03s UTC, and 0.16 for the flash registered on 20 Oct. 2012 at 20h48m28s UTC. These values together with the input data employed to obtain p from Eqs. (15) to (17) are summarized in Table 2. This analysis reveals that the strongest link that can be established corresponds to the flash imaged on 26 July 2012, since the probability of association with the α-Capricornid stream is of about 82%. The much lower probability obtained for the rest of the events in Table 2 (which is below 27% at best), implies that a sporadic origin should be assumed for these flashes.

### 3.2. Luminous efficiency

In the previous section we have employed a value of $\eta = 2\cdot10^{-3}$. We have followed the procedure described in Yanagisawa et al. (2006) to estimate the luminous efficiency for the α-Capricornid event and to check the validity of this assumption. Thus, the number N of expected impact flashes brighter than this CAP event was calculated by employing the formula given by Bellot Rubio et al. (2000):

$$N = F(m_o)\Delta t \left( \frac{2f\pi R^2}{\eta m_o V^2} E_d \right)^{1-s} A \quad (20)$$

where $F(m_o)$ is the flux or meteoroids with mass higher than $m_o$ ($2.3\cdot10^{-3}$ meteors km$^{-2}$ h$^{-1}$ according to the data collected by our meteor observing stations), $\Delta t$ is the observing time (4.7 hours in this case), A is the projected area of the observed lunar surface perpendicular to the α-Capricornid stream ($2.3\cdot10^{6}$ km$^2$ as determined from our images), and $E_d$ is the time-integrated optical energy flux of the flash observed on Earth ($9.7\cdot10^{-14}$ J m$^{-2}$). We have assumed f = 2. Besides, from Table 2 we have $m_o = 1.9\cdot10^{-6}$ kg, V = 31 km s$^{-1}$ and r = 2.5 (which in turn implies that s = 2.0 according to Eq. (16)). Then Eq. (20) implies that N = 0.6 with $\eta = 2\cdot10^{-3}$. This is close to the observational result, but since it does not fit very well the observations, the luminous efficiency was recalculated from this equation by employing the experimental value N = 1. This yields $\eta = 3.4\cdot10^{-3}$, which is very close to the assumed value. Thus, when the probability parameter p is recalculated with this new efficiency, the calculation yields 0.81, which confirms the strong association of this flash with the α-Capricornids. It should be taken into account that the assumed value of $\eta = 2\cdot10^{-3}$ was obtained from the analysis of Leonid impact flashes (Ortiz et al. 2006). Since Leonid meteoroids have a higher geocentric velocity ($V_g$ = 71 km s$^{-1}$) this difference in η could be due to a likely dependence of the luminous efficiency on velocity. To confirm this, additional α-Capricornid impact flashes should be analyzed in order to obtain with more accuracy the luminous efficiency for CAP events.

### 3.3. Meteoroid mass and crater size

Once the likely origin of the flashes listed in Table 1 was established, we have calculated the meteoroid mass and the size of the crater produced by these lunar impacts. Again, for sporadics we have considered $\eta = 2 \cdot 10^{-3}$, while we have set $\eta = 3.4 \cdot 10^{-3}$ for the impact flash associated to the α-Capricornids. The resulting values are listed in Table 3. In particular, the mass inferred for the CAP meteoroid (50 g) correlates fairly well with the observed masses (between 10 and 115 g) for fireballs produced on Earth by the largest meteoroids recorded by our meteor observing stations also in 2012 (Madiedo et al. 2014c). To calculate meteoroid diameters, spherical shape was assumed. Besides, an average bulk density of 1.8 g cm$^{-3}$ and 2.1 g cm$^{-3}$ was considered for sporadics and α-Capricornids, respectively (Babadzhanov and Kokhirova 2009). These sizes range between 3.6 and 6.8 cm.

To estimate crater sizes, an average impact angle of 45º was assumed for sporadic meteoroids. The calculation was performed by using the following crater-scaling equation for the Moon given by Gault, which is valid for craters with a diameter of up to about 100 meters in loose soil or regolith (Gault 1974, Melosh 1989):

$$D = 0.25 \rho_p^{1/6} \rho_t^{-0.5} E^{0.29} (\sin\theta)^{1/3} \quad (21)$$

In this relationship, where parameter values must be entered in mks units, D is the crater diameter, E is the kinetic energy of the impactor, $\rho_p$ and $\rho_t$ are the impactor and target bulk densities, respectively, and θ is the impact angle with respect to the horizontal. For the target bulk density we have $\rho_t$=1.6 g cm$^{-3}$. As shown in Table 3, the calculated crater sizes ranged between 0.6 and 1.0 m. Despite these being too small to be observed with instruments on Earth, these craters could be imaged by a probe orbiting the Moon, such as for instance the Lunar Reconnaissance Orbiter (LRO).

## 5 CONCLUSIONS

By using small telescopes we are performing a systematic monitoring of the night side of the Moon visible from the Earth in order to detect flashes produced during the collision of meteoroids on the lunar surface. In the context of this survey we have developed the MIDAS software, which can be employed to identify and analyze these impact flashes. In particular, we have defined a weighted filter to perform a temporal smoothing of the images in order to improve the detectability of fainter flashes. The software also obtains parameters such as the selenographic coordinates of these events and their apparent magnitude. This in turn provides, among other data, the impact kinetic energy, the impactor mass, and the impact angle.

We have also defined a parameter that measures the probability that a given flash is associated to a meteoroid stream. This definition takes into account the different sources contributing to the flux of impactors on the Moon, including meteoroid streams and the sporadic background. This probability parameter is based on the fundamental fact that only those meteoroids with a kinetic energy above the minimum detectable energy are included in the computations. According to the value of this parameter we have analyzed a series of lunar impact flashes recorded during 2011 and 2012. Despite some of them were recorded within the activity period of different meteor showers (namely, the δ-Leonids, the Virginids, the α-Capricornids, and the Orionids), this parameter has shown that all of them should be assigned to sporadic sources with high confidence, despite the impact geometry was compatible with these meteoroid streams. Only one of these events could be linked to the α-Capricornid stream, since the probability of such association was about 82 %. The masses for the meteoroids producing these flashes ranged between 48 and 302 g, with crater sizes between 0.6 and 1.0 m. In particular, we found a good correlation between the 50 g mass inferred for the α-Capricornid meteoroid and the masses calculated for the largest meteoroids producing fireballs in the Earth's atmosphere in 2012 (which ranged between 10 and 115 g). By analyzing this α-Capricornid flash, we have estimated the luminous efficiency for meteoroids from this stream impacting the Moon. According to our computations, this efficiency for CAP meteoroids would be of about $3.4\cdot10^{-3}$, which is close to the $2\cdot10^{-3}$ luminous efficiency value calculated for Leonid impact flashes. Nevertheless, additional observations of α-Capricornid impact flashes would be necessary to determine if these differences are related to a likely dependence of the luminous efficiency on meteoroid velocity.

# TABLES

| Date and time (UTC) | Selenographic coordinates | Duration (s) | Apparent magnitude | Stream |
|---|---|---|---|---|
| 9 Apr. 2011 20h38m08s | Lat: 24.4º N Lon: 64.2º W | 0.08 | 8.0 | SPO |
| 9 Apr. 2011 20h52m44s | Lat: 26.7º S Lon: 45.0º W | 0.04 | 8.5 | SPO |
| 11 Apr. 2011 0h05m06s | Lat: 12.4º S Lon: 55.9º W | 0.04 | 8.2 | SPO |
| 30 Dec. 2011 21h00m30s | Lat: 12.8º N Lon: 28.4º W | 0.04 | 8.5 | SPO |
| 26 Feb. 2012 21h40m10s | Lat: 23.3º S Lon: 28.6º W | 0.04 | 8.8 | DLE? |
| 28 Feb. 2012 23h05m16s | Lat: 31.6º N Lon: 35.3º W | 0.04 | 8.1 | SPO |
| 27 Mar. 2012 20h47m16s | Lat: 24.4º S Lon: 69.6º W | 0.06 | 9.8 | VIR? |
| 26 Jul. 2012 21h35m04s | Lat: 7.8º S Lon: 68.6º W | 0.16 | 8.7 | CAP? |
| 20 Oct. 2012 20h05m03s | Lat: 14.4º N Lon: 77.4º W | 0.04 | 8.0 | ORI? |
| 20 Oct. 2012 20h48m28s | Lat: 4.5º N Lon: 21.3º W | 0.08 | 8.6 | ORI? |

Table 1. Characteristics of the impact flashes discussed in the text.

| Date and time (UTC) | φ (º) | $ZHR^{ST}_{Earth}$ ($h^{-1}$) | r | $m_o$x$10^{-6}$ (kg) | $V_g$ (km $s^{-1}$) | V (km $s^{-1}$) | $E_m$x$10^6$ (J) | $\nu_{SPO}$ x$10^{-5}$ | ν x$10^{-4}$ | p | Stream |
|---|---|---|---|---|---|---|---|---|---|---|---|
| 26 Feb.2012 21h40m10s | 39 | 3 | 3.0 | 1.0 | 29 | 27 | 3.85 | 2.6 | 0.2 | 0.13 | DLE |
| 27 Mar.2012 20h47m16s | 34 | 5 | 3.0 | 0.87 | 30 | 34 | 3.80 | 2.6 | 0.2 | 0.27 | VIR |
| 26 Jul.2012 21h35m04s | 30 | 6 | 2.5 | 1.9 | 25 | 31 | 3.05 | 3.4 | 3.1 | 0.82 | CAP |
| 20 Oct.2012 20h05m03s | 25 | 25 | 2.9 | 0.028 | 67 | 65 | 3.15 | 3.3 | 0.03 | 0.19 | ORI |
| 20 Oct.2012 20h48m28s | 45 | 25 | 2.9 | 0.028 | 67 | 65 | 3.15 | 3.3 | 0.03 | 0.16 | ORI |

Table 2. Parameters in Eqs. 15, 16 and 17 employed to validate the meteoroid source of impact flashes.

| Date and time (UTC) | Stream | M (g) | $D_p$ (cm) | D (m) |
|---|---|---|---|---|
| 9 Apr. 2011, 20h38m08s | SPO | 302 | 6.8 | 1.0 |
| 9 Apr. 2011, 20h52m44s | SPO | 95 | 4.6 | 0.7 |
| 11 Apr. 2011, 0h05m06s | SPO | 121 | 5.0 | 0.8 |
| 30 Dec. 2011, 21h00m30s | SPO | 105 | 4.8 | 0.7 |
| 26 Feb. 2012, 21h40m10s | SPO | 80 | 4.4 | 0.7 |
| 28 Feb. 2012, 23h05m16s | SPO | 155 | 5.5 | 0.8 |
| 27 Mar. 2012, 20h47m16s | SPO | 48 | 3.7 | 0.6 |
| 26 Jul. 2012, 21h35m04s | CAP | 50 | 3.6 | 0.9 |
| 20 Oct. 2012 ,20h05m03s | SPO | 137 | 5.2 | 0.8 |
| 20 Oct. 2012, 20h48m28s | SPO | 157 | 5.5 | 0.8 |

Table 3. Meteoroid and impact crater parameters (M: meteoroid mass; $D_p$: meteoroid diameter; D: crater diameter).

# FIGURES

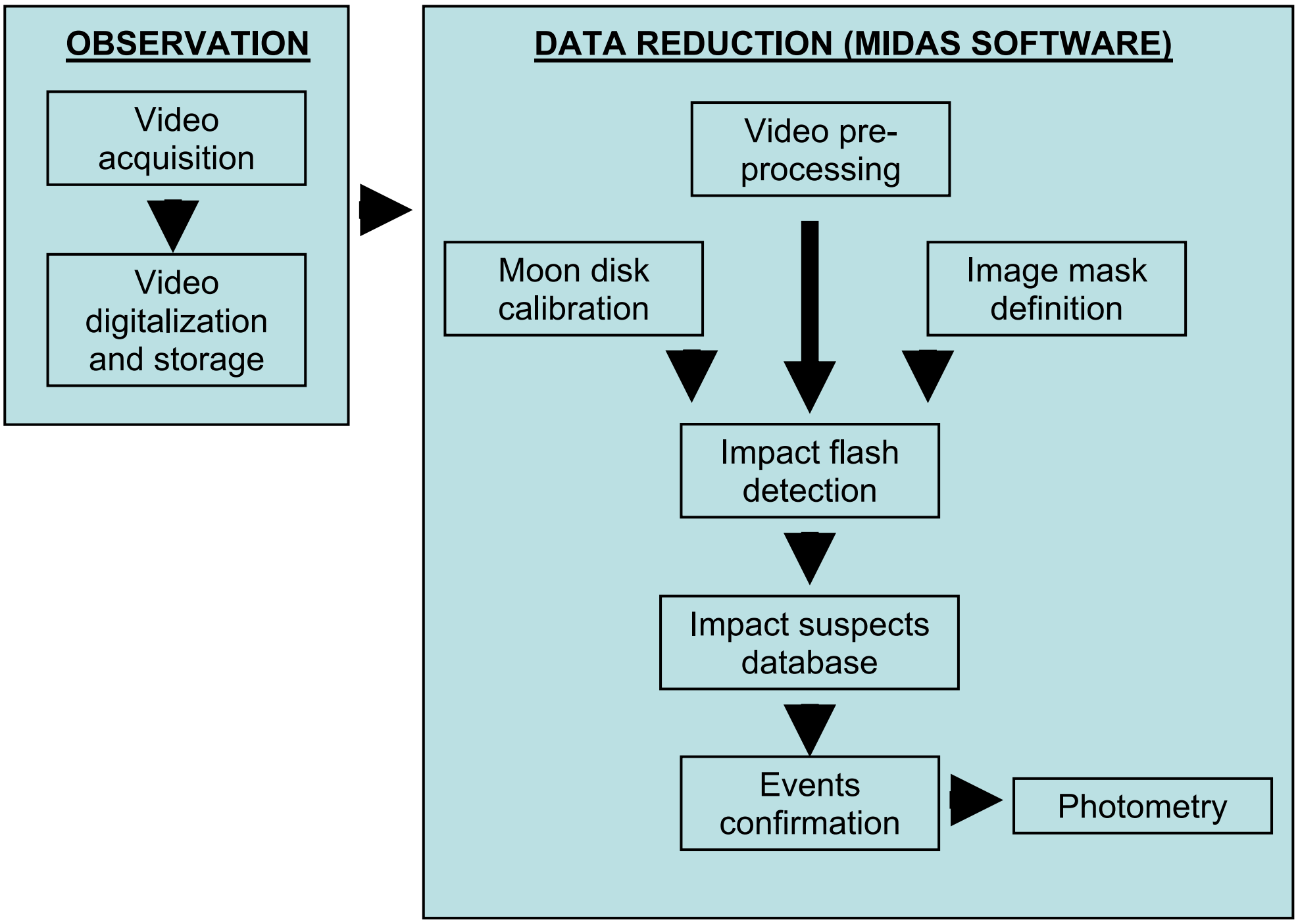

Figure 1. Summary of the lunar impact flash detection and analysis process.

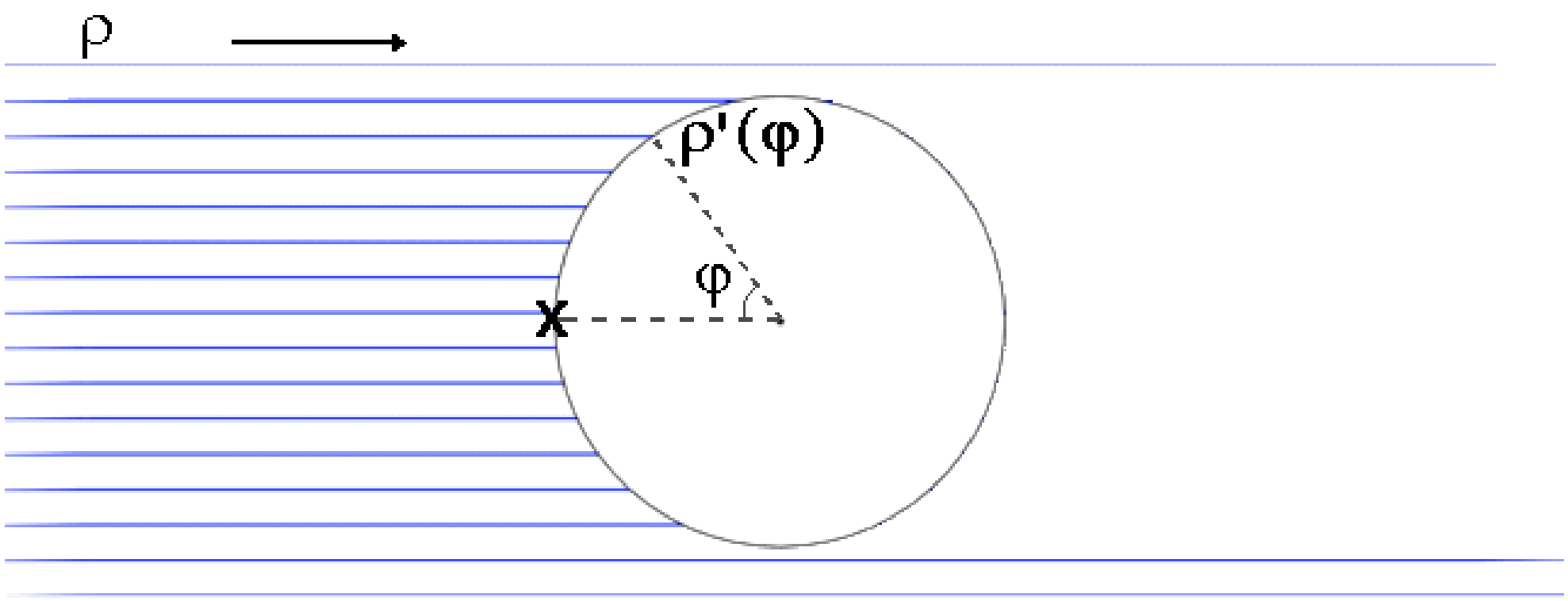

Figure 2. Schematic view of a meteoric tube with a homogeneous density ρ of meteoroids impacting on the lunar surface. The subradiant position is marked with a cross. The curvature of the lunar surface results in a position-dependent density of impacts ρ'(φ) on the Moon.

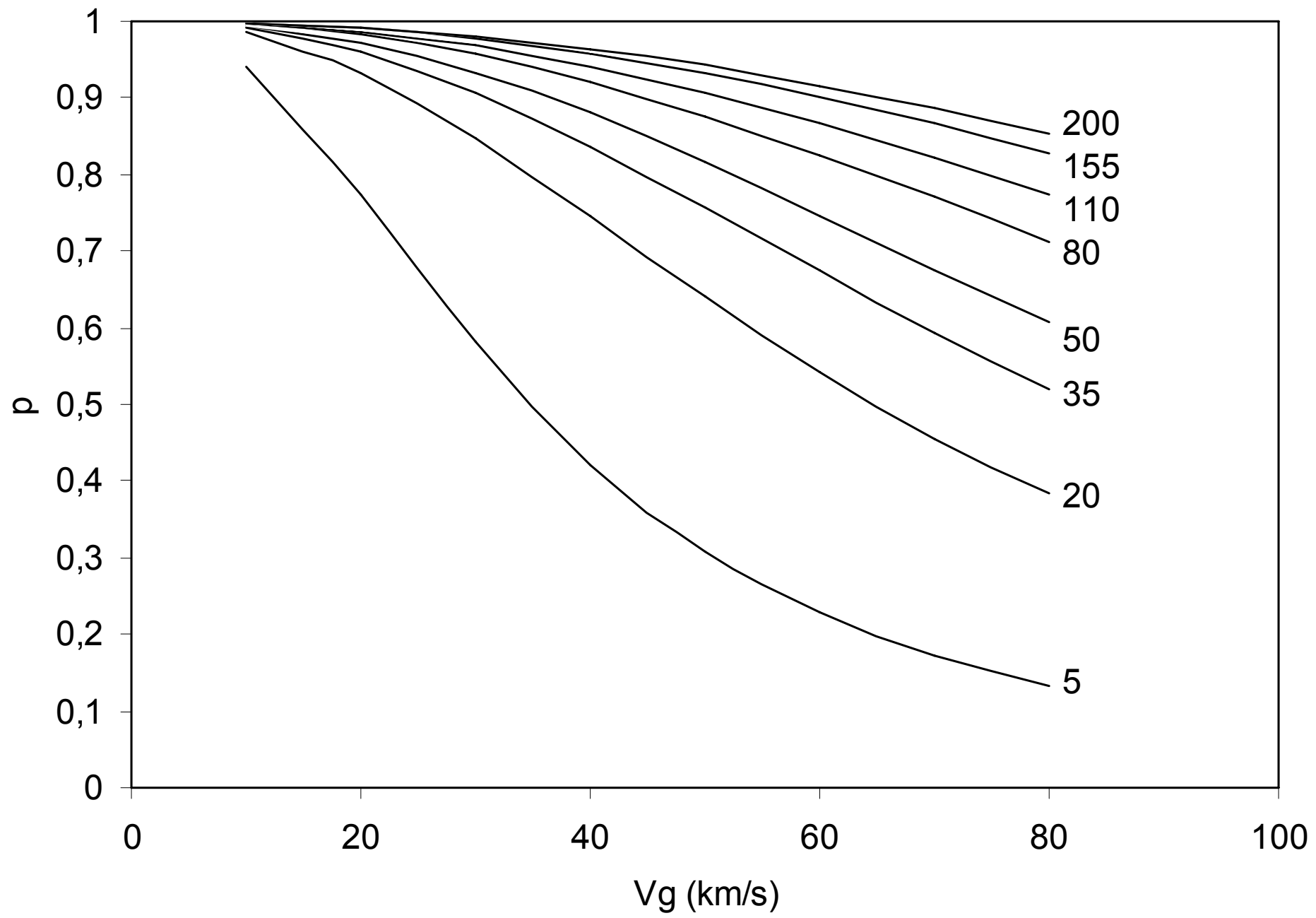

Figure 3. Impact probability predicted by Eq. (17) vs. meteoroid geocentric velocity for different ZHR values (in meteors $h^{-1}$). The computations were performed by considering r = 2.5, $\varphi = 45°$, $\sigma = 1$, and $\eta = 2{\cdot}10^{-3}$.

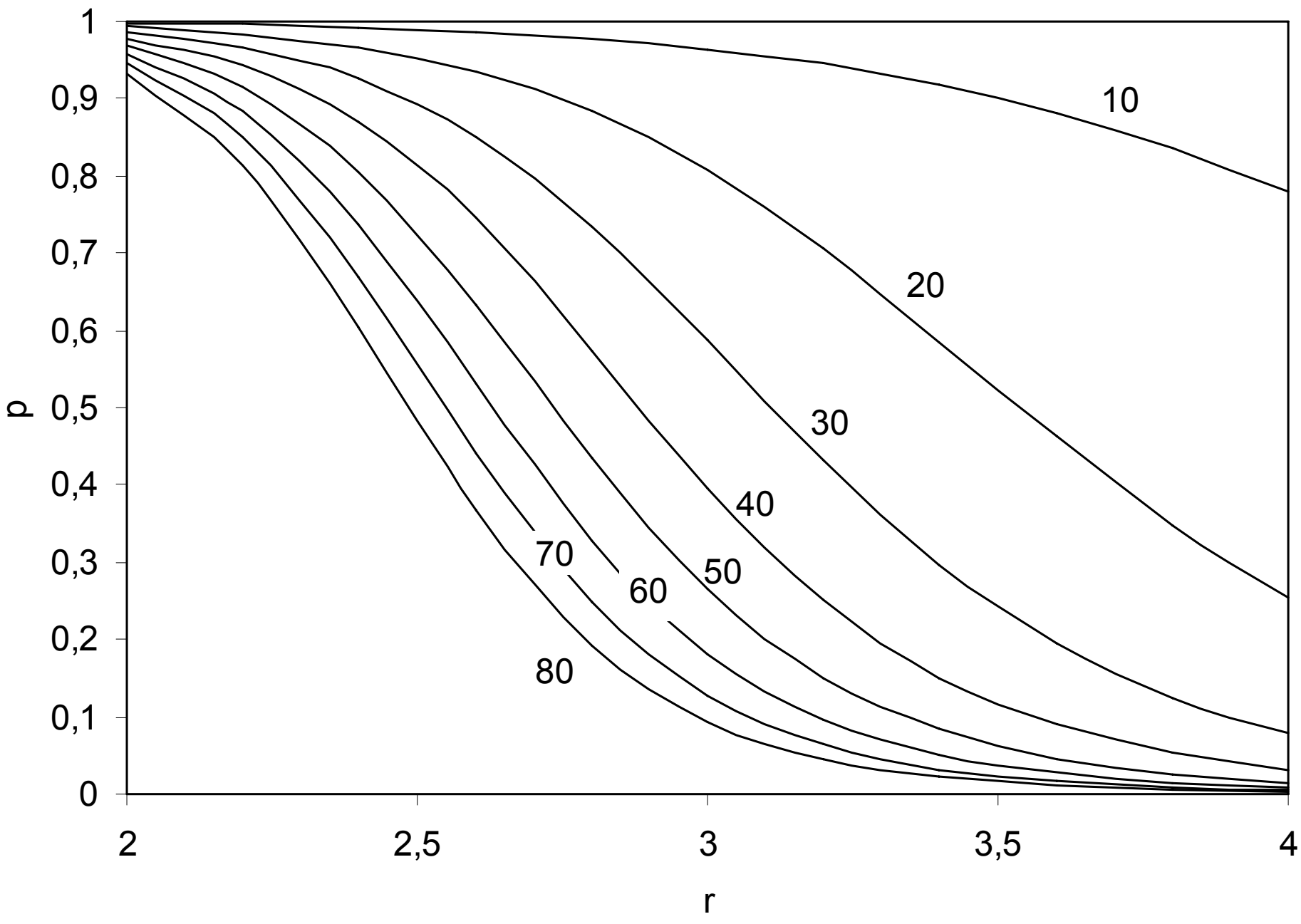

Figure 4. Impact probability predicted by Eq. (17) vs. stream population index for different Vg values (in km $s^{-1}$). The computations were performed by considering ZHR = 30, $\varphi = 45^{o}$, $\sigma = 1$, and $\eta = 2 \cdot 10^{-3}$.

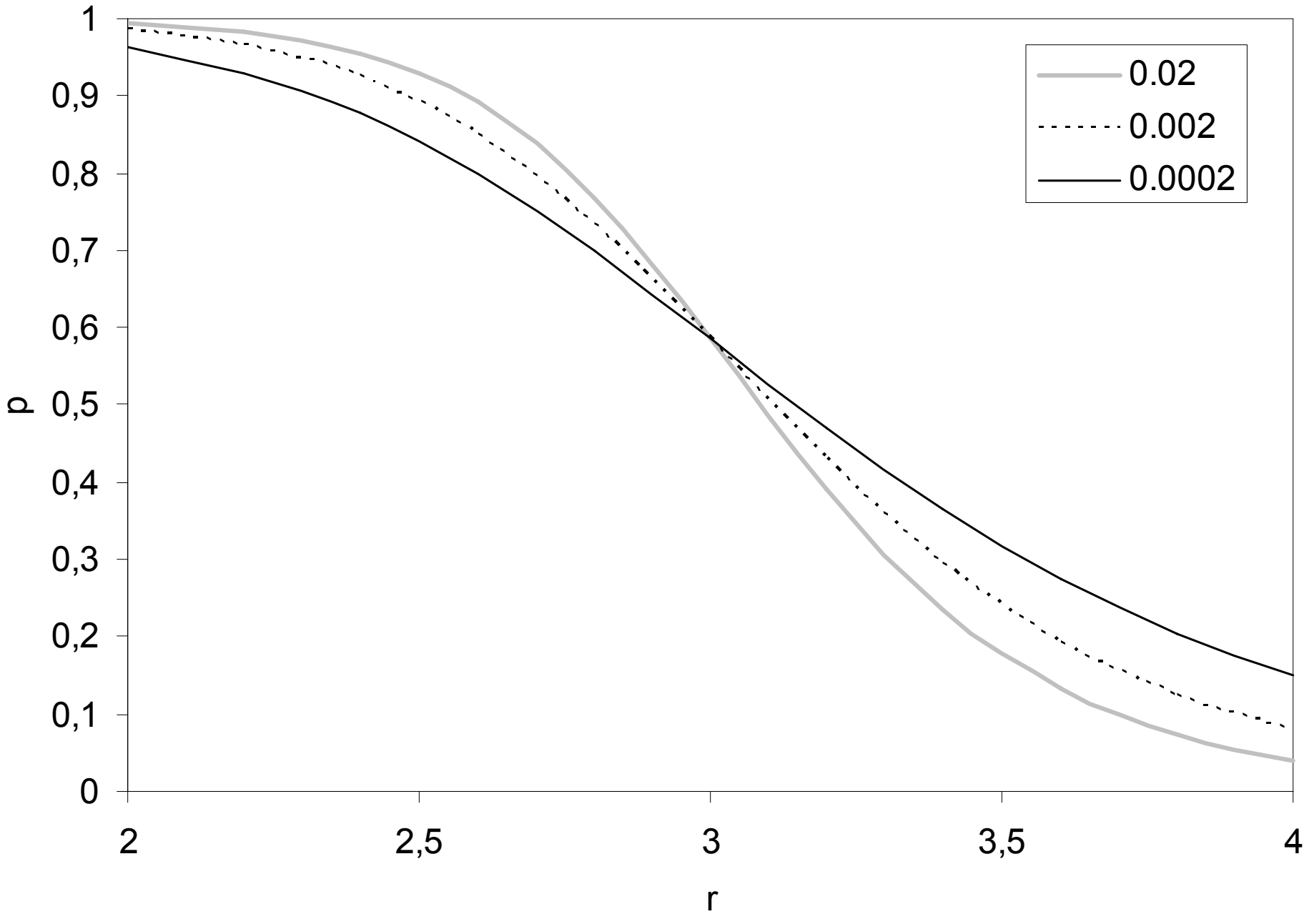

Figure 5. Impact probability predicted by Eq. (17) vs. stream population index for different luminous efficiency values. The computations were performed by considering ZHR = 30 and Vg = 30 km $s^{-1}$, $\varphi = 45º$, and $\sigma = 1$.

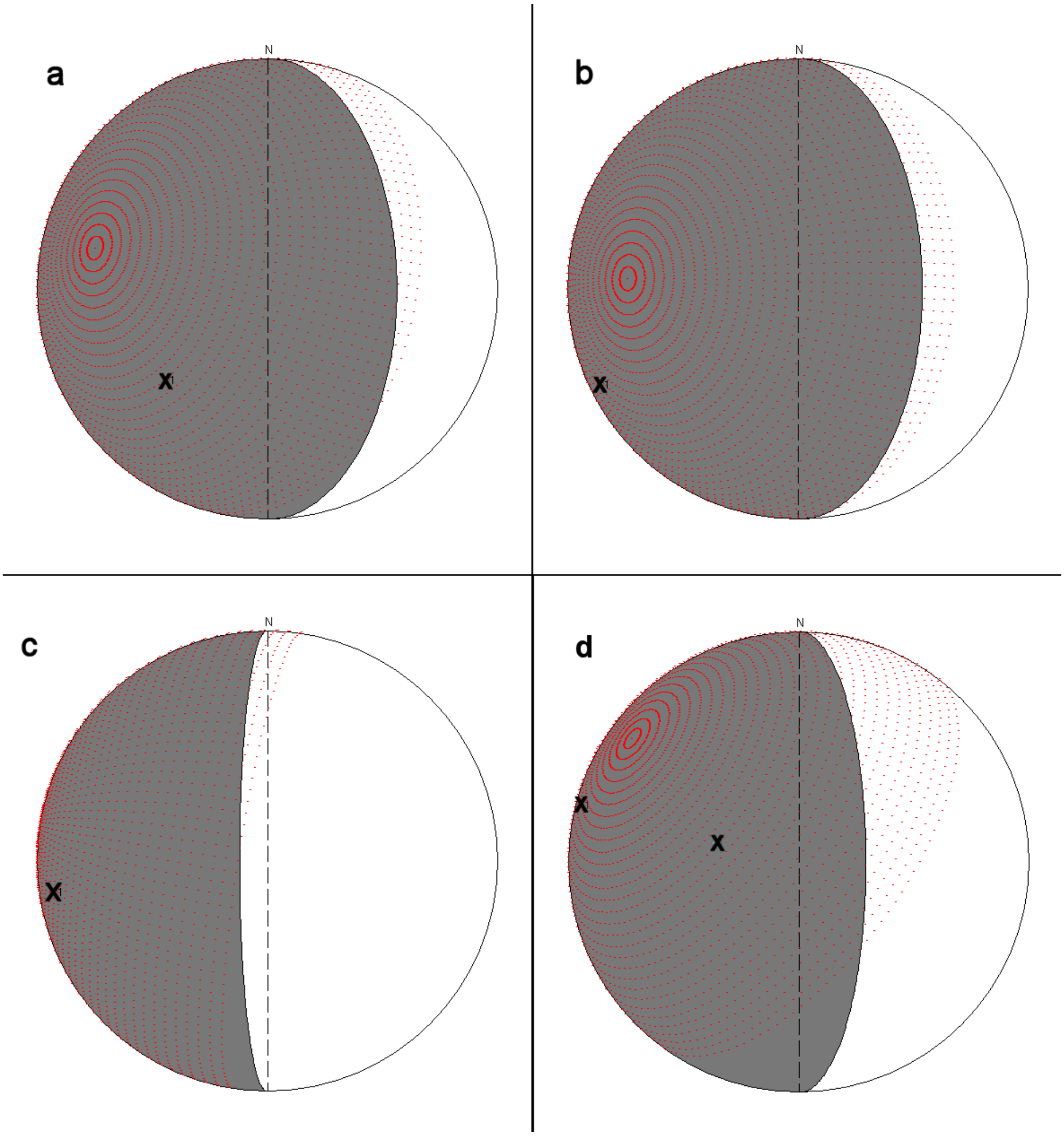

Figure 6. Impact geometry for the δ-Leonids (a), Virginids (b), α-Capricornids (c) and Orionids (d) on 26 Feb. 2012, 27 March 2012, 26 July 2012 and 20 October 2012, respectively. The impact position for flashes included in Table 2 is also shown. White region: area illuminated by the Sun; gray region: night side as seen from Earth; dotted region: area where meteoroids of the corresponding stream could impact.